\documentclass[a4paper]{torque2026}

\usepackage{graphicx}
\usepackage{lipsum}
\usepackage{amsmath}
\usepackage{amssymb}
\usepackage{graphicx}
\usepackage{svg}
\usepackage{url}
\usepackage{algorithm}
\usepackage{algpseudocode}
\usepackage{todonotes}
\begin{document}

\title{Accelerating Reinforcement Learning for Wind Farm Control via Expert Demonstrations}

\author{Marcus Binder Nilsen$^{1}$, Julian Quick$^{1}$, Tuhfe Göçmen$^{1}$, Nikolay Dimitrov$^{1}$ and Pierre-Elouan Réthoré$^{1}$}

\affil{$^1$Department of Wind and Energy Systems, Technical University of Denmark, Roskilde, Denmark}
% or synonym \affil{}

\ead{manils@dtu.org}
% or synonym \email{}

\begin{abstract}
Reinforcement learning (RL) offers a promising approach for adaptive wind farm flow control, yet its practical deployment is hindered by slow training convergence and poor initial performance, factors that could translate to years of reduced power output if an untrained agent were deployed directly. This work investigates whether domain knowledge from steady-state wake models can accelerate RL training and improve initial controller performance. We propose a pretraining methodology in which expert demonstrations are generated by deploying a PyWake-based steady-state optimizer within a dynamic wake simulation (WindGym), then used to initialize both the actor and critic networks of a Soft Actor–Critic agent via behavior cloning. Experiments on a 2×2 wind farm show that pretraining eliminates the costly initial learning phase: while an untrained agent underperforms the greedy zero-yaw baseline by approximately 12\%, pretraining raises initial performance to near-baseline levels. During online fine-tuning, all configurations converge within 250,000 environment steps to achieve similar performance, ultimately exceeding that of a lookup-table controller, which reaches approximately 7\% power gain after 500,000 steps. %An ablation study on dataset size reveals diminishing returns beyond 50 episodes, suggesting that excessive pretraining may over-constrain the policy to steady-state behavior. These findings demonstrate that modest amounts of steady-state expertise can substantially de-risk RL deployment for wind farm control without limiting long-term learning potential.
\end{abstract}

\section{Introduction}

The control of wind farms has been recognized as one of the ``grand challenges'' of modern wind energy systems \cite{veers2019grand,meyers2022wind}, particularly focusing on the mitigation of wake effects, which lead to decreased power production and negatively impact the structural integrity of downstream turbines \cite{Howland2020}.
Wind farm (flow) control (WFFC) has the potential to increase energy capture, decrease structural loads, thereby reducing operational costs and extending turbine lifetime~\cite{meyers2022wind}. However, WFFC remains a challenging optimization problem, primarily due to the inherent complexity of aerodynamic interactions between turbines, high-dimensional decision spaces, and significant uncertainties in environmental conditions.

% Reinforcement Learning (RL) as a Promising Solution
One possible solution is reinforcement learning (RL), which has recently emerged as a promising approach for adaptive data-driven WFFC \cite{abkar2023reinforcement, Gocmen2024}. RL methods are particularly attractive because of their capability to continually learn directly from interaction with dynamic environments and adapt accordingly, making them well-suited for managing the stochastic nature of wind conditions. %Recent research has demonstrated the potential of RL to outperform traditional methods in maximizing energy yield and reducing structural loads \cite{liew2023model}.

% Current Limitations of RL in Wind Farm Control
Nevertheless, the practical application of RL in WFFC faces significant challenges, notably slow training convergence and high computational demands resulting from RL’s inherent sample inefficiency, which often requires millions of interactions to converge \cite{duan2016benchmarkingdeepreinforcementlearning}. 

Because RL relies on experience gathered through environmental interaction, it inevitably exhibits an early learning phase, with performance below that of existing control strategies. For a real wind farm, such a phase could mean years of reduced power output and substantial financial loss, making direct deployment infeasible.

% Opportunity: Leveraging Existing Domain Knowledge
Fortunately, extensive domain knowledge already exists in the form of analytical wake models and steady-state optimization methods, which provide efficient and accurate benchmarks for optimal turbine operation under simplified conditions. This motivates the central research question of this work:

% Research Objective and Contribution (Clear Statement)
% Brief Overview of Proposed Method
\begin{itemize}
\item \textbf{Can knowledge from steady-state models improve the initial and overall performance of RL agents for wind-farm control?}
\end{itemize}

% To address this question, we propose a novel pretraining methodology that integrates insights from steady-state optimization into the RL training process. By transferring such domain knowledge, we aim to enhance the agent’s sample efficiency, accelerate convergence, and ultimately improve the practicality of RL-based wind-farm control.
Several works have explored how steady-state models can integrate within the RL framework. Zhao et al.\ \cite{8999726} proposed a knowledge-assisted 'Deep Deterministic Policy Gradient' framework that uses analytical models for safety constraints and reward shaping during online learning, though without pretraining the policy networks. Stanfel et al.\ \citep{9147946} demonstrated that Q-tables trained offline, based on steady-state optimization using FLORIS \cite{floris}, can transfer to quasi-dynamic simulations. Most closely related, Bizon Monroc et al.\ \cite{Bizon_Monroc_2024} derived deterministic reference policies from lookup tables to initialize actor-critic networks. In contrast, we propose a pretraining methodology based on expert demonstrations collected during a dynamic wake simulation, training both the policy and value networks of a Soft Actor-Critic (SAC) agent. Furthermore, we analyze the effect of pretraining dataset size on learning performance—an aspect that has not been previously investigated. By transferring such domain knowledge, we aim to enhance the agent's initial performance and ultimately improve the practicality of RL-based wind-farm control.

In this work, a dynamic wake simulation serves as a surrogate for a real wind farm, allowing us to evaluate pretraining benefits in a controlled setting. While field deployment would raise additional questions—such as when to update the policy—one could envision a multi-stage pipeline in which the agent is first pretrained on a steady-state model, refined in a dynamic simulator, and finally fine-tuned on the real plant.

% The rest of the paper is organized as follows. Section~\ref{sec:methodology} describes the simulation environment and the methodology for the expert dataset generation, supervised pretraining, and fine-tuning. Section~\ref{sec:results} presents results, comparing “pre-training” against training without prior knowledge, and analyzes how dataset size affects learning. Finally, Section~\ref{sec:conclusion} concludes with a discussion of limitations and future work. % Intro and some kind of stat of the art -> 1 page
\section{Methodology}
\label{sec:methodology}

% \subsection{Reinforcement Learning Background}

RL formalizes sequential decision-making as a Markov Decision Process (MDP), defined by the tuple $(\mathcal{S}, \mathcal{A}, P, r )$:
\begin{itemize}
    \item $\mathcal{S}$: the state space, represented as a vector of per-turbine features including local wind speed and direction, yaw angle, and their 25-step averages. All values are normalized to $[-1, 1]$.
    \item $\mathcal{A}$: the action space specifying yaw adjustments for each turbine. An action of 0 keeps the yaw unchanged, while $\pm 1$ corresponds to the maximum allowable change per timestep. The action is defined as: 
    $\gamma'_i = \gamma_i + a_{t,i} \cdot \gamma_{\max}$   \\
    Here $\gamma_i$ and $\gamma'_i$ is the current and previous yaw angle of turbine $i$, and $\gamma_{\max}$ is the maximum allowed yaw step. This work uses a maximum yaw rate of $0.5 \deg / \sec$ 
    \item $P(s_{t+1} \mid s_t, a_t)$: the transition model, assumed deterministic as turbines perfectly follow control commands.
    \item $r(s_t, a_t)$: the reward, defined as the relative power gain compared to a baseline controller that aligns all turbines with the inflow:
    \begin{equation}
        r_t = \left( \frac{\sum_{i=1}^{N_\text{turb}} P_{i,\text{agent}}(t)}{\sum_{i=1}^{N_\text{turb}} P_{i,\text{baseline}}(t)} - 1 \right)
    \end{equation}
    
    % \item $\gamma \in [0,1)$: the discount factor, set to 0.99.
\end{itemize}

At each time step $t$, the agent observes a state vector $s_t \in \mathcal{S}$ and selects an action vector $a_t \in \mathcal{A}$, where subscript $i$ denotes the per-turbine component (e.g., $a_{t,i}$).

The RL objective is to find a stochastic policy $\pi_\psi(a \mid s)$ that maximizes the expected return:

\begin{equation}
J(\psi) = \mathbb{E}_{\tau \sim \pi_\psi} \left[ \sum_{t=0}^T \lambda^t r(s_t, a_t) \right],
\end{equation}
where $\tau = (s_0, a_0, r_0\ldots, s_T,a_T,r_T)$ is a trajectory induced by the policy and environment, and $\lambda$ is used to denote the discount factor, set to 0.99 in this work.

Actor–critic algorithms are commonly used to solve this optimization problem, where the actor $\pi_\psi(a \mid s)$ samples actions and the critic $V_\phi(s)$ estimates expected returns. This framework is particularly effective for continuous, high-dimensional control problems such as WFFC.

This work uses the \texttt{WindGym} environment \cite{WindGym} and the Soft Actor–Critic (SAC) algorithm \cite{haarnoja2018softactorcriticoffpolicymaximum} for the policy, as described below.

\subsection{WindGym Environment} 

\texttt{WindGym} \cite{WindGym} is a custom, open-source RL environment for wind-farm flow control. The environment utilizes DYNAMIKS \cite{Dynamiks, wesJulia}, which is a collection of different fidelity simulation tools. This work uses dynamic wake meandering (DWM) \cite{larsen2007dynamic} to simulate the transient evolution of the wakes in a turbulent inflow. 
The simulated wind farm consists of a $2 \times 2$ grid of DTU 10MW turbines \cite{DTU10MW}, each represented with their performance ($C_P$) and thrust ($C_T$) coefficients using PyWake \cite{pywake}. Turbines are spaced $5D$ apart in both streamwise and spanwise directions, where $D$ is the rotor diameter.

Each \texttt{WindGym} episode corresponds to a fixed inflow scenario (wind speed, direction, turbulence intensity) and runs until the flow has passed through the farm 20 times. The simulation time step is 5 s, while the agent issues control commands every 10 s. During each episode, the policy accumulates experience by observing turbine-level measurements and adjusting yaw angles accordingly.

\subsection{Soft Actor-Critic (SAC)}

This work uses the SAC algorithm \cite{haarnoja2018softactorcriticoffpolicymaximum}, a state-of-the-art algorithm for continuous control. SAC maximizes a trade-off between expected return and policy entropy. This encourages exploration by changing the optimization problem to:
\begin{equation}
J(\pi) = \mathbb{E}_{\pi} \left[ \sum_{t=0}^T \lambda^t \big( r[s_t, a_t] + \alpha \mathcal{H}[\pi(\cdot|s_t)] \big) \right],
\end{equation}
where $\alpha$ is the temperature parameter controlling the exploration–exploitation balance, and $\mathcal{H}(\pi(\cdot|s_t))$ is the policy entropy.

SAC employs an actor–critic architecture with the following components:

\textbf{Actor network ($\pi_\psi$):}  
Outputs a Gaussian distribution over actions. The network maps states $s$ to a mean $\mu_\psi(s)$ and log standard deviation $\log \sigma_\psi(s)$ in the unconstrained action space. Actions are then sampled and passed through a $\tanh$ squashing function. For full details, please refer to the original paper \cite{haarnoja2018softactorcriticoffpolicymaximum}:
% \begin{equation}
% a = \text{scale} \cdot \tanh(\hat{a}) + \text{bias}, \quad \hat{a} \sim \mathcal{N}(\mu_\theta(s), \sigma_\theta(s)).
% \end{equation}

\textbf{Critic networks ($Q_{\phi}, Q_{\phi_2}$):}  
Two independent critics estimate state–action values $Q(s, a)$ to mitigate positive bias in value estimates. % Each critic is trained to minimize the Bellman error using target networks and the clipped double-Q trick.

\textbf{Network architecture:}  
All networks employ a standard multilayer perceptron (MLP) structure with two hidden layers of 256 units and ReLU activations. 

SAC was chosen for its robustness, stability, and high sample efficiency in continuous control, and has demonstrated strong performance in domains such as robotics and wind-farm control \cite{neustroev2022deep}. For a more detailed explanation of the algorithm, we refer to the original paper \cite{haarnoja2018softactorcriticoffpolicymaximum}.

\subsection{Expert Policy from Steady-State Optimization}
\label{subsec:expert}

To accelerate RL learning, expert demonstrations are generated using a steady-state optimization method based on analytical wake models. The expert policy is constructed using PyWake \cite{pywake} and the \texttt{SerialRefine} optimizer \cite{Fleming_2022}.
Given free-stream wind conditions $(U_\infty, \theta, I)$, the expert computes optimal yaw angles $a^\text{expert} \in \mathbb{R}^{N_\text{turb}}$ that maximize steady-state farm power output, assuming perfect knowledge of the inflow. This results in a high-performing, though idealized, policy.

The expert is then deployed in \texttt{WindGym} to collect realistic trajectories:
\begin{equation}
\tau = \big[ (s_t, a_t^\text{expert}, r_t^\text{expert}) \big]_{t=0}^T,
\end{equation}
Each represents the temporal evolution of expert decisions in turbulent conditions. 

To analyze the impact of expert demonstration volume on the agent's initial performance and training stability, this study compares RL agents initialized with varying levels of data derived from PyWake optimization. Three datasets are used for comparison, each with a different number of randomly sampled episodes, with wind conditions sampled like described in section \ref{subsec:traineval}. The naming convention and dataset sizes are presented in Table \ref{tab:dataset}. Note that we also tested a larger number of episodes in datasets, but this showed no meaningful difference. 

\begin{table}[H]
\centering
\caption{Dataset name and size}
\vspace{10pt}
\label{tab:dataset}
\begin{tabular}{l||c|c|c|c}
\textbf{Dataset name}                  & None & Small & Medium & Large \\ \hline
\textbf{Episodes in dataset} & 0    & 10     & 50    & 200 
\end{tabular}
\end{table}

\subsection{Pretraining with Expert Demonstrations}
The expert trajectories are used for offline pretraining of both the actor and critic networks before online RL training begins. The goal is to improve the agent's initial performance by exposing it to expert examples of effective control behavior.

\subsubsection{Actor Pretraining}

The actor is trained by maximizing the log-likelihood of expert actions under the current policy.

The expert dataset $\mathcal{D}_\text{expert} = \{(s_i, a_i)\}_{i=1}^N$ is unpacked into a flat collection of state–action pairs, and split into an 80-20 split for training and validation.

The SAC uses a squashed Gaussian policy, which must be inverted to compute the likelihood in the unconstrained space:
\begin{equation}
\hat{a}_i = \text{atanh}\left( \frac{a_i - \text{bias}}{\text{scale}} \right).
\end{equation}
The log-likelihood of $\hat{a}_i$ under the Gaussian $\mathcal{N}(\mu_\psi(s_i), \sigma_\psi(s_i))$ is corrected with the change-of-variable term:
\begin{equation}
\log p(a_i \mid s_i) = \log \mathcal{N}(\hat{a}_i \mid \mu_\psi(s_i), \sigma_\psi(s_i)) - \sum_d \log\left(1 - \tanh^2(\hat{a}_{i,d})\right) + \log |\text{scale}|.
\end{equation}
The behavior cloning loss is then:
\begin{equation}
\label{eq:loss}
\mathcal{L}_\text{BC}(\psi) = -\frac{1}{N_B} \sum_{j=1}^{N_B} \log p(a_j \mid s_j).
\end{equation}

The pretraining is done for up to 200 epochs using the Adam optimizer ($\text{lr} = 3 \times 10^{-5}$), with early stopping based on validation loss (patience = 5 epochs). The best-performing parameters are restored at the end. The full algorithm can be seen in algorithm \ref{alg:bc_actor}.

\begin{algorithm}[t]
\caption{Behavior Cloning Pretraining of SAC Actor}
\label{alg:bc_actor}
\begin{algorithmic}[1]
\Require Expert dataset $\mathcal{D}_{\text{expert}} = \{(s_i, a_i)\}_{i=1}^N$, initial actor parameters $\psi$
\State Split $\mathcal{D}_{\text{expert}}$ into $\mathcal{D}_{\text{train}}$ and $\mathcal{D}_{\text{val}}$ (80/20)
\State Initialize: best loss $L_{\text{best}} \gets \infty$, patience counter $p \gets 0$
\For{epoch $= 1$ to $E_{\max}$}
    \State Shuffle $\mathcal{D}_{\text{train}}$
    \For{each minibatch $\mathcal{B} \subset \mathcal{D}_{\text{train}}$}
        \State Compute loss $\mathcal{L}_{\text{BC}}(\psi)$ using Eq. \ref{eq:loss}
        \State Update actor: $\psi \gets \psi - \eta \nabla_\psi \mathcal{L}_{\text{BC}}$
    \EndFor
    \State Evaluate validation loss $\mathcal{L}_{\text{val}}$ on $\mathcal{D}_{\text{val}}$
    \If{$\mathcal{L}_{\text{val}} < L_{\text{best}}$}
        \State $L_{\text{best}} \gets \mathcal{L}_{\text{val}}$
        \State Save current $\psi$ as $\psi^*$
        \State $p \gets 0$ \Comment{Reset patience counter}
    \Else
        \State $p \gets p + 1$
    \EndIf
    \If{$p \geq P_{\text{early}}$}
        \State \textbf{break} \Comment{Early stopping triggered}
    \EndIf
\EndFor
\State \Return Pretrained actor parameters $\psi^*$
\end{algorithmic}
\end{algorithm}

\subsubsection{Critic Pretraining}

The critic network is trained to predict the discounted returns from the expert trajectories.

For each expert trajectory $\tau = \{(s_t, a_t, r_t)\}_{t=0}^T$, the discounted returns are calculated as:
\begin{equation}
G_t = \sum_{k=t}^T \lambda^{k-t} r_k,
\end{equation}
The value function $V_\phi(s)$ is then fitted to minimize the mean squared error:
\begin{equation}
\mathcal{L}_\text{VC}(\phi) = \frac{1}{N_B} \sum_{j=1}^{N_B} \left( V_\phi(s_j) - G_j \right)^2.
\end{equation}

To stabilize training, target returns are normalized to zero mean and unit variance. The critic pretraining uses the Adam optimizer with a learning rate of $1 \times 10^{-3}$, batch size 64, and the same early stopping criteria as those for the actor network.

\subsection{Training and Evaluation Protocol}
\label{subsec:traineval}

After pretraining, the SAC agent is fine-tuned via standard RL updates for $1\times10^{6}$ environment steps. Training is divided into discrete episodes, each corresponding to a fixed wind inflow scenario and turbulence field. At the start of each episode, wind speed is sampled from $U_\infty \sim \mathcal{U}(8, 15)$~m/s, wind direction from $\theta \sim \mathcal{U}(255^\circ, 285^\circ)$, and initial yaw angles from $\gamma_i \sim \mathcal{U}(-15^\circ, 15^\circ)$, while turbulence intensity is fixed at $I = 0.05$. Note that $\mathcal{U}$ denotes a uniform distribution. Turbulence boxes are sampled uniformly from a pre-generated library of $N_\text{box} = 10$ fields. These conditions ensure significant wake steering potential while including scenarios without wake losses. Because wind speed and direction are part of the agent's observation space, the policy learns to generalize across the full operating envelope; the reported training steps therefore reflect learning over all sampled conditions jointly, not convergence for a single inflow scenario. Training is repeated across $N_\text{seed} = 5$ random seeds to reduce initialization sensitivity, with agent weights saved every $2.5\times10^{5}$ steps for intermediate evaluation.

\paragraph{Evaluation procedure.}
All policies are evaluated on 6 new unseen turbulence realizations, across a discrete grid of wind conditions:
\[
\theta \in \{255^\circ, 260^\circ, \ldots, 285^\circ\}, \quad U_\infty \in \{8, 10, 12, 13\}~\text{m/s}.
\]
Each evaluation runs for $T_{\text{eval}} = 3600$s, corresponding to one hour.

% Most RL algorithms incorporate some form of stochasticity to ensure sufficient exploration, such as SAC, which learns a Gaussian distribution. However, that means that the algorithm can be evaluated without this stochasticity. This work employs stochastic action sampling, rather than relying on the mean of the output distribution. This matches the sampling procedure used during training and better reflects real-world execution.

% All results are reported as mean $\pm$ standard deviation across the 5 SAC seeds.

In addition to SAC, we evaluate the following baselines:

\begin{itemize}
    \item \textbf{Greedy (Zero-Yaw)}: All turbines remain aligned with the incoming flow ($\gamma = 0$). This serves as a simple, realistic, and also in line with the standard WFFC optimisation practices.
    
    \item \textbf{PyWake}: Uses a look-up table (LUT) of optimal yaw angles generated by PyWake and SerialRefine, but infers inflow conditions solely from the most upstream turbine. To mitigate measurement noise, the inferred wind speed and direction are Polyak-averaged over time:
    \[
    \hat{x}_t = (1 - \rho) \hat{x}_{t-1} + \rho x_t, \quad \text{with } \rho = 0.05.
    \]
\end{itemize}

The mean power gain over all the conditions can then be found according to equation \ref{eq:mean}. 

\begin{equation}
\label{eq:mean}
   \dfrac{1}{n_{\text{wd}}n_{\text{ws}}n_{\text{box}}n_{\text{seed}}T} \sum_{i_{\text{wd}}}^{n_{\text{wd}}} \sum_{i_{\text{ws}}}^{n_{\text{ws}}} \sum_{i_{\text{box}}}^{n_{\text{box}}}\sum_{i_{\text{seed}}}^{n_{\text{seed}}}\sum_{t}^{T} P_{\text{farm}}(t)
\end{equation}

Each SAC configuration is trained with one of four pretraining levels (None, Small, Medium, Large) and five random seeds, for a total of 20 training runs. During training, the model is saved every $2.5\times10^{5}$ training steps. Each snapshot is evaluated over a grid of wind conditions comprising 4 wind speeds, 7 wind directions, and 6 turbulence boxes. Thus, each snapshot is tested on $4\times 7\times 6 = 168$ cases, and each training run on $5\times 168 = 840$ total cases. % 3 pages
\section{Results and discussion}
\label{sec:results}

\begin{figure}[H]
    \centering
    \includegraphics[width=.9\linewidth]{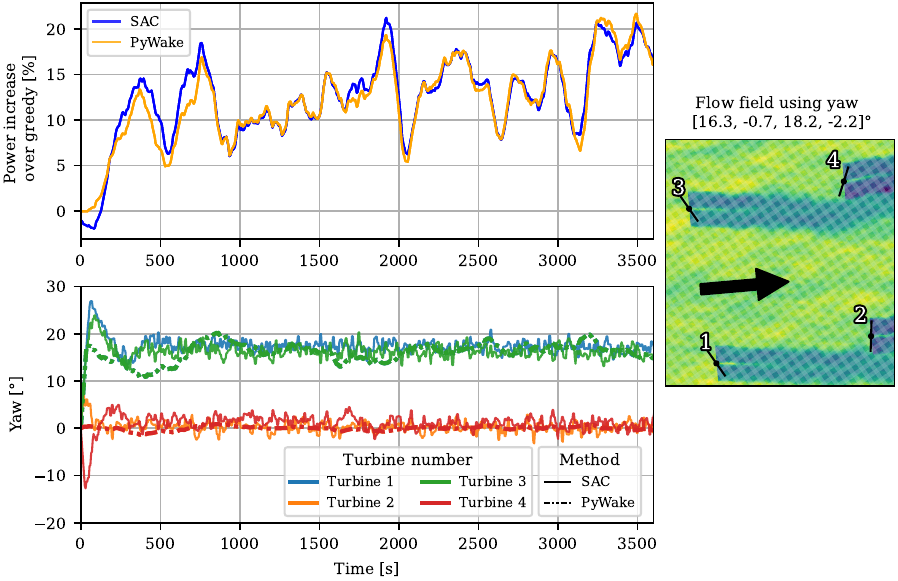}
    \caption{Single-case evaluation at 12\,m/s and $275^\circ$. Top: wind farm power gain relative to Greedy for SAC and PyWake. Bottom: yaw control actions per turbine; the PyWake yaw for the two upstream and two downstream turbines coincide. Right: snapshot of the flow field at $t=3000$\,s with the applied yaw angles.}
    \label{fig:single_case}
\end{figure}

Figure \ref{fig:single_case} shows a representative evaluation at $U_\infty=12$\,m/s and $\theta=275^\circ$, a slightly misaligned inflow case. Both SAC and the PyWake LUT controller outperform the Greedy baseline over most of the horizon, with SAC yielding the highest total farm power for this case. The corresponding yaw trajectories indicate broadly similar control patterns between the two controllers, with PyWake exhibiting symmetry in the front and back rows due to its steady-state structure.

\begin{figure}[H]
    \centering
    \includegraphics[width=.8\linewidth]{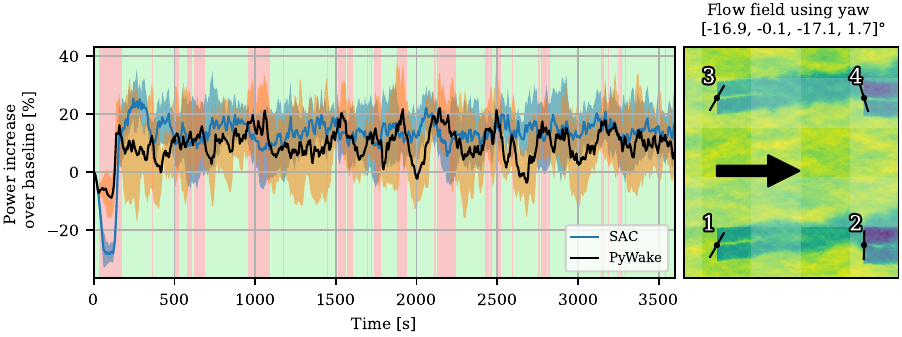}
     \caption{Comparison of SAC and PyWake controllers for $\theta = 270^\circ$ and $U_\infty = 10$\,m/s. 
     Left: mean wind farm power gain (\%) relative to the Greedy baseline; shaded regions indicate 1 standard deviation across turbulence boxes (and seeds for SAC). 
     Right: flow snapshot for one turbulence realization. 
     Green and background indicate higher mean power for SAC and PyWake, respectively.}
    
    \label{fig:single_sweep}
\end{figure}

\vspace{-0.5cm}
Figure \ref{fig:single_sweep} compares the fully trained SAC (Large pretraining) to PyWake at $U_\infty=10$\,m/s and $\theta=270^\circ$, a fully aligned inflow case. Curves show the mean and 1 standard deviation across turbulence realizations, where SAC aggregates over $5$ seeds $\times$ $6$ boxes (60 cases) and PyWake over 6 boxes. Background shading indicates which controller attains higher mean power at each time (green for SAC, red for PyWake) and it shows that SAC overperforms the PyWake controller majority of the time for this inflow case, although absolute differences are relatively small.

\begin{figure}[H]
    \centering
    \includegraphics[width=.9\linewidth]{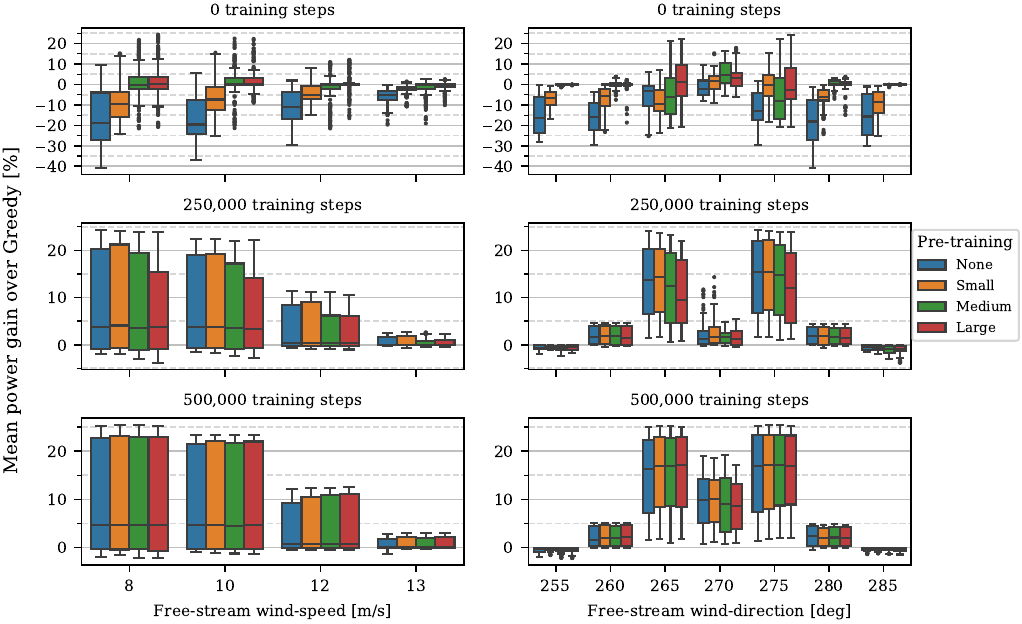}
    \caption{Box plots of wind farm power gain (\%) relative to Greedy across wind speeds (left) and wind directions (right) after 0, $2.5\times 10^{5}$, and $5\times 10^{5}$ training steps. Pretraining most strongly improves the untrained policy and low-wind regimes; little gain is observed at edge directions where wake interactions are weak.}
    \label{fig:boxplots}
\end{figure}

Figure \ref{fig:boxplots} summarizes the distribution of percentage power gain over the Greedy case across wind speeds (left column) and directions (right column) at three training milestones (0 vs. 250,000 vs. 500,000 steps). Pretraining yields large benefits at initialization (top row), which diminish as online training proceeds. This is most clear at low wind speeds, where wake steering has the largest potential impact. Power gains are negligible for edge directions ($255^\circ$ and $285^\circ$), where turbines are largely unwaked.

\begin{figure}[H]
    \centering
    \includegraphics[]{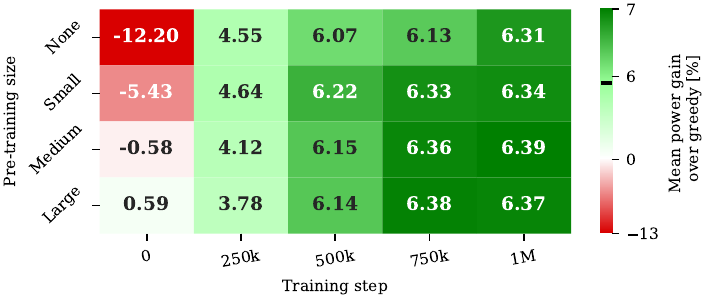}
    \caption{Heatmap of the mean percentage increase in power for all pre-training sizes and training steps. Note that the PyWake agent has a mean increase of $5.96 \%$. This value is indicated by the black line in the colorbar on the right}
    \label{fig:heatmap}
\end{figure}

To compare controllers with a single scalar per snapshot and pretraining level, we compute the mean percentage gain over the full wind grid and turbulence boxes as in Eq.~\eqref{eq:mean} (Section~\ref{sec:methodology}). Figure \ref{fig:heatmap} shows that the SAC agent without pretraining initially underperforms Greedy by $\sim 12\%$, whereas pretraining substantially raises the initial performance. By 250,000 steps, all SAC variants reach a mean gain of roughly $4\%$. After 500,000 steps, SAC surpasses the PyWake LUT controller slightly, exceeding a mean gain of $6\%$ (PyWake mean is $5.96\%$; indicated by the black tick in the color bar right-hand side of Figure~\ref{fig:heatmap}).

Several observations from Figure~\ref{fig:heatmap} merit discussion. The convergence of all pretraining variants to similar performance by 250,000 steps suggests that pretraining is not necessary for achieving good performance, at least in this small farm case. However, its practical value lies in avoiding the costly initial learning phase, where an untrained agent would have a $12\%$ power loss relative to the normal Greedy performance.

A counterintuitive pattern emerges at intermediate training steps: the Small dataset outperforms Medium and Large at 250,000 steps ($4.64\%$ vs.\ $3.78\%$). This could be caused by the model overcommitting to the expert's steady-state behavior for the larger dataset, thereby failing to explore and discover the optimal policy. However, note that it could also be due to a relatively low number of agent seeds used.

% We hypothesize that larger pretraining datasets overcommit the policy to the expert's steady-state behavior, thereby impeding the discovery of dynamic strategies that deviate from the static optimum. Since the expert assumes perfect inflow knowledge and ignores temporal dynamics, heavier pretraining may encode biases requiring additional experience to unlearn. A smaller dataset provides sufficient signal to escape poor initial behaviors while preserving exploratory flexibility.

Regarding practical implications, generating expert demonstrations and pretraining incurs a computational cost; however, the results show diminishing returns for larger datasets at intermediate training budgets. Since all configurations converge by 500,000 steps, the dataset size primarily affects the transient phase of the algorithm. A modest dataset of 50 episodes offers the best trade-off, capturing most initialization benefit without over-constraining adaptation. % 3.5 pages
\section{Conclusions and Further Work}
\label{sec:conclusion}

This study demonstrates that behavior-cloning pretraining with expert trajectories significantly improves the early performance of RL-based wind farm control without constraining long-term learning. Immediately after deployment, the untrained agent produced roughly 12\% less power than the greedy zero-yaw baseline; however, pretraining with 50–200 expert episodes raised this starting point to approximately equal performance. During online fine-tuning, all agents converged within approximately 250,000 environment interactions to around 4\% mean gain over the greedy, ultimately surpassing the PyWake-based lookup strategy by surpassing 6\% power gain after 500,000 steps. Crucially, pretraining exhibited no observable downside: it significantly improved initial performance while achieving identical levels of final power gain.

Several directions merit further exploration. Expert demonstrations could be sampled more strategically—prioritizing informative wind conditions rather than uniform sampling—to convey more knowledge per episode. The current agent is not penalized for excessive yaw motion; incorporating yaw-rate penalties would yield more fatigue-aware control. Finally, future studies should assess scalability on larger wind farms, where richer wake interactions may amplify the benefits of informed pretraining.

% Half a page

\ack
The authors gratefully acknowledge the computational and data resources provided on the Sophia HPC Cluster at the Technical University of Denmark, DOI: 10.57940/FAFC-6M81.
\\
\vspace{-0.5cm}
\bibliographystyle{iopart-num}
% \bibliography{Torque26_references}
\bibliography{refs}

@article{Dynamiks,
        title={Dynamiks 0.0.4: An open-source Dynamic Wind System Simulator},
        author={Mads M. Pedersen and Julia Steiner and Marcus Binder Nilsen and Jonas Lohmann and Emily Louise Hodgson and Riccardo Riva and Niels Troldborg and S{\o}ren Juhl Andersen and Gunner Larsen and David Robert Verelst and Pierre-Elouan R\'{e}thor\'{e}},
        url="https://gitlab.windenergy.dtu.dk/DYNAMIKS/dynamiks",
        publisher={DTU Wind, Technical University of Denmark},
        year={2026},
        month={1}
}

@article{pywake,
    title={PyWake 2.5.0: An open-source wind farm simulation tool},
    author={Mads M. Pedersen and Alexander Meyer Forsting and Paul van der Laan and Riccardo Riva and Leonardo A. Alcayaga Romàn and Javier Criado Risco and Mikkel Friis-Møller and Julian Quick and Jens Peter Schøler Christiansen and Rafael Valotta Rodrigues and Bjarke Tobias Olsen and Pierre-Elouan Réthoré},
    url={https://gitlab.windenergy.dtu.dk/TOPFARM/PyWake},
    publisher={DTU Wind, Technical University of Denmark},
    year={2023},
    month={2}
}

@inproceedings{neustroev2022deep,
  title={Deep reinforcement learning for active wake control},
  author={Neustroev, Grigory and Andringa, Sytze PE and Verzijlbergh, Remco A and De Weerdt, Mathijs M},
  booktitle={Proceedings of the 21st International Conference on Autonomous Agents and Multiagent Systems},
  pages={944--953},
  year={2022}
}

@Article{floris,
    author = {NREL},
    title = {{FLORIS}. Version 4.2.1},
    year = {2024},
    publisher = {GitHub},
    journal = {GitHub repository},
    url = {https://github.com/NREL/floris}
}

@article{abkar2023reinforcement,
  title={Reinforcement learning for wind-farm flow control: Current state and future actions},
  author={Abkar, Mahdi and Zehtabiyan-Rezaie, Navid and Iosifidis, Alexandros},
  journal={Theoretical and Applied Mechanics Letters},
  pages={100475},
  year={2023},
  publisher={Elsevier}
}

@article{Gocmen2024,
  title={Data-driven wind farm flow control and challenges towards field implementation},
  author={G{\"o}{\c{c}}men, Tuhfe and Liew, Jaime and Kadoche, Elie and Dimitrov, Nikolay and Riva, Riccardo and Andersen, S{\o}ren Juhl and Lio, Alan W.H. and Quick, Julian and R{\'e}thor{\'e}, Pierre-Elouan and Dykes, Katherine},
  journal={Renewable and Sustainable Energy Reviews},
  note={Under Review},
  year={2024}
}

@article{Howland2020,
    author = "Howland, Michael F. and Dabiri, John O.",
    doi = "10.3390/en14010052",
    title = "Influence of Wake Model Superposition and Secondary Steering on Model-Based Wake Steering Control with {SCADA} Data Assimilation",
    journal = "Energies",
    year = "2020"
}

@article{veers2019grand,
  title={Grand challenges in the science of wind energy},
  author={Veers, Paul and Dykes, Katherine and Lantz, Eric and Barth, Stephan and Bottasso, Carlo L and Carlson, Ola and Clifton, Andrew and Green, Johney and Green, Peter and Holttinen, Hannele and others},
  journal={Science},
  volume={366},
  number={6464},
  pages={eaau2027},
  year={2019},
  publisher={American Association for the Advancement of Science}
}

@article{meyers2022wind,
  title={Wind farm flow control: prospects and challenges},
  author={Meyers, Johan and Bottasso, Carlo and Dykes, Katherine and Fleming, Paul and Gebraad, Pieter and Giebel, Gregor and G{\"o}{\c{c}}men, Tuhfe and Van Wingerden, Jan-Willem},
  journal={Wind Energy Science Discussions},
  volume={2022},
  pages={1--56},
  year={2022},
  publisher={G{\"o}ttingen, Germany}
}

@article{larsen2007dynamic,
    title = {Dynamic wake meandering modeling},
    author = {Larsen, Gunner C and Aagaard Madsen, H and Bing{\"o}l, Ferhat},
    year = {2007}
}

@misc{DTU10MW,
    title = {The {DTU} {10-MW} Reference Wind Turbine},
    author = {Christian Bak and Frederik Zahle and Robert Bitsche and Taeseong Kim and Anders Yde and Henriksen, {Lars Christian} and Hansen, {Morten Hartvig} and Blasques, {Jos{\'e} Pedro Albergaria Amaral} and Mac Gaunaa and Anand Natarajan},
    year = {2013},
    language = {English},
    note = {Danish Wind Power Research 2013; Conference date: 27-05-2013 Through 28-05-2013},
}

@misc{duan2016benchmarkingdeepreinforcementlearning,
      title={Benchmarking Deep Reinforcement Learning for Continuous Control}, 
      author={Yan Duan and Xi Chen and Rein Houthooft and John Schulman and Pieter Abbeel},
      year={2016},
      eprint={1604.06778},
      archivePrefix={arXiv},
      primaryClass={cs.LG},
      url={https://arxiv.org/abs/1604.06778}, 
}

@misc{WindGym,
  title        = {WindGym},
  author       = {DTU},
  year         = 2025,
  note         = {\url{
      Available: https://github.com/DTUWindEnergy/WindGym 
      } [Accessed: 27-05-2025]}
}

@article{Fleming_2022,
doi = {10.1088/1742-6596/2265/3/032109},
url = {https://dx.doi.org/10.1088/1742-6596/2265/3/032109},
year = {2022},
month = {may},
publisher = {IOP Publishing},
volume = {2265},
number = {3},
pages = {032109},
author = {Fleming, Paul A. and Stanley, Andrew P. J. and Bay, Christopher J. and King, Jennifer and Simley, Eric and Doekemeijer, Bart M. and Mudafort, Rafael},
title = {Serial-Refine Method for Fast Wake-Steering Yaw Optimization},
journal = {Journal of Physics: Conference Series}
}

@INPROCEEDINGS{9147946,
  author={Stanfel, Paul and Johnson, Kathryn and Bay, Christopher J. and King, Jennifer},
  booktitle={2020 American Control Conference (ACC)}, 
  title={A Distributed Reinforcement Learning Yaw Control Approach for Wind Farm Energy Capture Maximization}, 
  year={2020},
  volume={},
  number={},
  pages={4065-4070},
  doi={10.23919/ACC45564.2020.9147946}}

@ARTICLE{8999726,
  author={Zhao, Huan and Zhao, Junhua and Qiu, Jing and Liang, Gaoqi and Dong, Zhao Yang},
  journal={IEEE Transactions on Industrial Informatics}, 
  title={Cooperative Wind Farm Control With Deep Reinforcement Learning and Knowledge-Assisted Learning}, 
  year={2020},
  volume={16},
  number={11},
  pages={6912-6921},
  doi={10.1109/TII.2020.2974037}}

@article{Bizon_Monroc_2024,
doi = {10.1088/1742-6596/2767/3/032017},
url = {https://doi.org/10.1088/1742-6596/2767/3/032017},
year = {2024},
month = {jun},
publisher = {IOP Publishing},
volume = {2767},
number = {3},
pages = {032017},
author = {Bizon Monroc, C and Bušić, A and Dubuc, D and Zhu, J},
title = {Towards fine tuning wake steering policies in the field: an imitation-based approach},
journal = {Journal of Physics: Conference Series},
}

@misc{haarnoja2018softactorcriticoffpolicymaximum,
      title={Soft Actor-Critic: Off-Policy Maximum Entropy Deep Reinforcement Learning with a Stochastic Actor}, 
      author={Tuomas Haarnoja and Aurick Zhou and Pieter Abbeel and Sergey Levine},
      year={2018},
      eprint={1801.01290},
      archivePrefix={arXiv},
      primaryClass={cs.LG},
      url={https://arxiv.org/abs/1801.01290}, 
}

@Article{wesJulia,
    AUTHOR = {Steiner, J. and Hodgson, E. L. and van der Laan, M. P. and others},
    TITLE = {A multi-fidelity model benchmark for wake steering of a large turbine in a neutral {ABL}},
    JOURNAL = {Wind Energy Science Discussions},
    VOLUME = {2025},
    YEAR = {2025},
    PAGES = {1--32},
    URL = {https://wes.copernicus.org/preprints/wes-2025-200/},
    DOI = {10.5194/wes-2025-200}
}

\end{document}